\documentclass[12pt]{article}
\usepackage{amssymb,amsmath,epsfig}

\begin{document}

\title{\bf Axially Symmetric Shear-free Fluids in $f(R,T)$ Gravity}

\author{Ifra Noureen \thanks{ifra.noureen@gmail.com; ifra.noureen@umt.edu.pk} ${}^{(a)}$, M. Zubair
\thanks{mzubairkk@gmail.com; drmzubair@ciitlahore.edu.pk} ${}^{(b)}$ \\
${}^{(a)}$ Department of Mathematics,\\University of Management and
Technology, Lahore, Pakistan.\\ ${}^{(b)}$ Department of
Mathematics,\\ COMSATS Institute of Information Technology, Lahore,
Pakistan.}
\date{}
\maketitle

\begin{abstract}

In this work we have discussed the implications of shear-free
condition on axially symmetric anisotropic gravitating objects in
$f(R,T)$ theory. Restricted axial symmetry ignoring rotation and
reflection enteries is taken into account for establishment of
instability range. Implementation of linear perturbation on
constitutive modified dynamical equations yield evolution equation.
This equation associates adiabatic index $\Gamma$ with material and
dark source components defining stable and unstable regions in
Newtonian (N) and post-Newtonian (pN) approximations.

\end{abstract}


\section{Introduction}

In a recent work \cite{1}, we have discussed the instability range
of axial system in $f(R, T)$ gravity, $R$ being Ricci scalar and $T$
denoting trace of energy momentum. In continuation to \cite{1}, we
plan to study the evolution of axially symmetric sources evolving
under shear-free condition. Shear tensor has significant importance
in the of structure formation, inhomogeneity factors and
relativistic stellar phenomenons. Many authors \cite{2}-\cite{8}
have discussed the impact of shear tensor in gravitational evolution
and consequences of its vanishing value. It has been established
recently \cite{9} that the vanishing shear case has equivalence with
the homology conditions and so has enormous significance in
astrophysics \cite{10, 11}.

Possible aberrancies from most commonly studied spherical symmetry
extended the concerns with non-spherical geometries of gravitating
systems. Recent developments suggest that incidental variations in
spherical symmetry prevails the more realistic scenarios. In this
work, we have considered restricted axially symmetric system
(ignoring reflection and rotation about symmetry axis) with
anisotropic matter configuration evolving under shear-free
condition. This study is conducted in $f(R, T)$ gravity by taking
into account a consistent $f(R,T)$ model (satisfying
$\frac{df}{dr}\geq0, \frac{d^{2}f}{dr^{2}}\geq0$).

Study of gravitational collapse and its end-state has gained great
deal of attention in past years. Dominance of inward acting gravity
force over outward acting pressure leads to the gravitational
collapse due to which star contracts to a point. A gravitating
system remains stable until pressure keeps on balancing the
gravitational pull and collapses when pressure to gravity balance
fluctuates. The range of stability/instability primarily depends on
mass, supersessive stars lose stability more rapidly. Besides mass
there are various factors that can modify instability range
considerably such as shear, isotropy, anisotropy, radiation etc.

Instability problem received enormous attention after the seminal
contribution of Chanderashekar \cite{12}, he set stability criterion
of gravitating sources in the form of adiabatic index $\Gamma$
associating variation in pressure with varying energy density. The
criterion for stability of gravitating systems with anisotropic
matter distribution is established in \cite{13}. Herrera and his
collaborators \cite{14}-\cite{18} contributed greatly to study the
dynamics of relativistic fluids to discuss the factors contributing
in gravitational evolution. They also presented the general
framework to deal with axially symmetric sources and their
applications \cite{19}-\cite{21}.

Gravitational interaction can be described in different manners, so
far the scheme of general theory of relativity (GR) being a
self-consistent approach has been largely used. It explains
gravitational phenomenons adequately at cosmological scales
\cite{22}. However, developments in observational situations
\cite{23}-\cite{27} suggests that GR can not be the only theory that
can be appropriate for all scales.

Many treatments have been made for suitable description of
gravitational theories on large scales and deal with the issue of
cosmic speed-up \cite{28}-\cite{37}, by developing alternative
gravitational theories \cite{38}-\cite{44} for e.g. $f(R)$,
$f(R,T)$, $f(G)$, $f(T,\mathcal{T})$, Brans-Dicke theory and so on.
Such alternative theories are dynamically equivalent to GR at
cosmological levels. Herein, we workout the gravitational evolution
in $f(R,T)$ theory introduced by Harko \cite{45}. In this theory the
matter content is considered to have interaction with the geometry.
Many authors \cite{46}-\cite{48} have discussed the cosmological and
thermodynamic implications in $f(R,T)$ gravity.

The purpose of this work is to explore the implication of shear-free
condition on dynamical instability of a restricted class of axially
symmetric anisotropic systems in $f(R,T)$ gravity. The gravitational
action in $f(R,T)$ includes an arbitrary function of $R$ and $T$,
written as \cite{45}
\begin{equation}\label{1}
\int dx^4\sqrt{-g}[\frac{f(R, T)}{16\pi G}+\mathcal{L} _ {(m)}],
\end{equation}
where $\mathcal{L} _ {(m)}$ is matter Lagrangian and $g$ is the
metric. A variety of choices of $\mathcal{L} _ {(m)}$ can be dealt
with, each of which represent a specific configuration of
relativistic matter.

The article is arranged as: Section \textbf{2} includes the
discussion of considered spacetime, matter configuration, kinematic
variables along with shear-free condition and modified dynamical
equations. In section \textbf{3}, we introduced the $f(R,T)$ model
and applied linear perturbation on conservation equations. Section
\textbf{4} contains the discussion of weak field limit. Summary of
the results is given in last section that is followed by an
appendix.

\section{Modified Dynamical Equations With Vanishing Shear}

In this section, we develop the modified field field equations in
framework of $f(R,T)$ gravity by considering axial system with
anisotropic matter evolving under shear-free condition.

\subsection{Metric and Matter Configuration}

The dynamical systems lacking spherical symmetry have received great
attention in recent past, since such abberations describes the more
realistic phases of gravitational evolution. The general line
element for axially symmetric gravitating sources constitutes five
metric functions (independent), given by
\begin{eqnarray}\nonumber
ds^2&=&-A^2(t, r, \theta)dt^{2}+B^2(t, r, \theta)dr^{2}+r^2B^2(t, r,
\theta) d\theta^{2} +C^2(t, r, \theta)d\phi^{2}\\\label{1'}&&+2G(t,
r, \theta)dtd\theta+2H(t, r, \theta)dtd\phi,
\end{eqnarray}
Herein, our goal is to study the implications of vanishing shear
scalar on collapsing phenomenon. We have restricted the character of
spacetime by imposing the highest degree of symmetry to avoid the
complexities generated from the terms of reflection and rotation
about symmetry axis. Thus, we have restricted the general line
element (\ref{1'}) to somehow handle the problem under consideration
and manipulate the results analytically. Considering vorticity-free
case i.e. ignoring $dtd\theta$ and $dtd\phi$ terms in (\ref{1'}),
the reduced form of line element containing three independent metric
functions takes following form
\begin{equation}\label{1''}
ds^2=-A^2(t,r,\theta)dt^{2}+B^2(t,r,\theta)(dr^{2}+r^2 d\theta^{2})
+C^2(t,r,\theta)d\phi^{2}.
\end{equation}

The matter distribution is taken to be anisotropic carrying unequal
pressure stresses. The energy momentum tensor for usual matter is
defined as \cite{19}
\begin{eqnarray}\nonumber&&
T^{(m)}_{uv}=(\rho+p_\perp)V_{u}V_{v}-(K_uK_v-\frac{1}{3}h_{uv})(P_{zz}-P_{xx})-
(L_uL_v-\frac{1}{3}h_{uv})(P_{zz}\\\label{3}&&-P_{yy})+Pg_{uv}+2K_{(u}L_{v)}P_{xy},
\end{eqnarray}
where
\begin{equation}\nonumber
P=\frac{1}{3}(P_{xx}+P_{yy}+P_{zz}), \quad h_{uv}=g_{uv}+V_{u}V_{v},
\end{equation}
where $P_{xx}, P_{yy}, P_{zz}$ and $P_{xy}$ denote corresponding
anisotropic pressure stresses, satisfying $P_{xy}=P_{yx}$ and
$P_{xx}\neq P_{yy}\neq P_{zz}$. The four vectors in radial and axial
directions are described by $K_u$ and $L_u$ respectively, while
$V_{u}$ stands for four-velocity and $\rho$ is energy density. The
above mentioned quantities are associated as
\begin{equation}\label{4}
V_{u}=-A\delta^{0}_{u},\quad K_{u}=B\delta^{1}_{u},\quad
L_{u}=rB\delta^{2}_{u}.
\end{equation}

\subsection{Kinematic Variables and Shear-free Condition}

Kinematic variables play significantly important role in the
description of self-gravitating sources. Herein three kinematic
variables are contributing in the systems evolution, i.e.,
acceleration $a_u$, expansion scalar $\Theta$ and the shear tensor
$\sigma_{uv}$.  These variables are obtained as follows
\begin{equation}\label{5}
a_{u}=V^{v}V_{u;v}=(0, \frac{A'}{A}, \frac{A^\theta}{A}, 0),
\end{equation}
\begin{equation}\label{6}
\Theta=V_{;u}^{u}=\frac{1}{A}\left(\frac{2\dot{B}}{B}+\frac{\dot{C}}{C}\right),
\end{equation}
\begin{equation}\label{7}
\sigma_{uv}=V_{(u;v)}-a_{(u}V_{v)}-\frac{1}{3}\Theta(h_{uv}).
\end{equation}
The non-zero components of $\sigma_{uv}$ are
\begin{eqnarray}\label{8}&&
\sigma_{11}=\frac{B^2}{3A}\left(\frac{\dot{B}}{B}-\frac{\dot{C}}{C}\right),\\\label{9}&&
\sigma_{22}=\frac{r^2B^2}{3A}\left(\frac{\dot{B}}{B}-\frac{\dot{C}}{C}\right),\\\label{9a}&&
\sigma_{33}=\frac{-2C^2}{3A}\left(\frac{\dot{B}}{B}-\frac{\dot{C}}{C}\right).
\end{eqnarray}
The expression for shear scalar is obtained as \cite{20}
\begin{equation}\label{9b}
\sigma=\frac{1}{A}\left(\frac{\dot{B}}{B}-\frac{\dot{C}}{C}\right).
\end{equation}
Here dot and prime indicate derivatives w.r.t time and radial
coordinates respectively. The shearing effects and inhomogeneity
factors within the collapsing system play an essential role in the
formation of the trapped surfaces and the apparent horizon. It is
remarked in \cite{6} that presence of shear causes distortion of
trapped surface geometry which strongly implicates the development
of naked singularities as end state of gravitational evolution. It
is worth mentioning here that absence of shearing effects in
collapsing cloud lead to the formation of a black hole. Thus,
absence or presence of shear have significant impression on
kinematics of the gravitating source and its evolution.

The gravitating source under consideration is assumed to have
vanishing shear, i.e., $\sigma=0$, that leads to
$\frac{\dot{B}}{B}=\frac{\dot{C}}{C}$. We shall employe shear-free
condition to study the dynamics of axially symmetric gravitating
source.

\subsection{Modified Field Equations}

As mentioned earlier in introduction that different choices for
matter Lagrangian can be taken to study the dynamics of a
gravitating system, each choice represents a particular set of field
equations. Here, we take $\mathcal{L} _ {(m)}=- \rho$, $8\pi G = 1$
and vary the extended gravitational action (\ref{1}) w.r.t metric
$g_{uv}$ as follows
\begin{eqnarray}\nonumber
G_{uv}&=&\frac{1}{f_R}\left[(f_T+1)T^{(m)}_{uv}+\rho g_{uv}f_T+
\frac{f-Rf_R}{2}g_{uv}\right.\\\label{9c}&+&\left.(\nabla_u\nabla_v-g_{uv}\Box)f_R\right],
\end{eqnarray}
where $\Box=\nabla^{u}\nabla_{v}$, $f_R\equiv df(R,T)/dR$,
$f_T\equiv df(R,T)/dT$, $\nabla_{u}$ stands for covariant derivative
and $T^{(m)}_{uv}$ denote the usual matter stress energy tensor.
Corresponding components of effective Einstein tensor on account of
shear-free condition are
\begin{eqnarray}\nonumber
G^{00}&=&\frac{1}{A^2f_R}\rho+\frac{1}{A^2f_R}\left[\frac{f-Rf_R}{2}
-\frac{3\dot f_R}{A^2}\frac{\dot{B}}{B}
-\frac{f_R'}{B^2}\left(\frac{1}{r}+\frac{2B'}{B}-\frac{C'}{C}\right)\right.\\\label{10}&&\left.
-\frac{f_R^\theta}{r^2B^2}\left(\frac{2B^\theta}{B}-\frac{C^\theta}{C}\right)
+\frac{f_R''}{B^2}\right],
\\\label{11}
G^{01}&=&\frac{-1}{A^2B^2f_R}\left[\frac{A'}{A}\dot{f_R}+\frac{\dot{B}}{B}f_R'-\dot{f_R}'
\right],
\\\label{12}
G^{02}&=&\frac{-1}{r^2A^2B^2f_R}\left[\frac{A^\theta}{A}\dot{f_R}+\frac{\dot{B}}{B}f_R^\theta-\dot{f_R}^\theta
\right],
\end{eqnarray}
\begin{eqnarray}\nonumber
G^{11}&=&\frac{1}{B^2f_R}\left[P_{xx}(f_T+1)+\rho f_T-\frac{\dot
f_R}{A^2} \frac{\dot{A}}{A}- \frac{f-Rf_R}{2}-\frac{
f_R^{\theta\theta}}{r^2B^2}- \frac{\ddot{f_R}}{A^2}
\right.\\\label{13}&&\left.+\frac{f_R'}{B^2}\left(\frac{1}{r}-\frac{A'}{A}+\frac{B'}{B}
-\frac{C'}{C}\right)+
\frac{f_R^\theta}{r^2B^2}\left(\frac{3B^\theta}{B}-\frac{A^\theta}{A}-\frac{C^\theta}{C}\right)\right],
\\\label{14}
G^{12}&=&\frac{1}{r^2B^4f_R}\left[P_{xy}(f_T+1)+f_R'^\theta-
\frac{B^\theta}{B}f'_R-\frac{B'}{B}f_R^\theta\right],
\\\nonumber
G^{22}&=&\frac{1}{r^2B^4f_R}\left[P_{yy}(f_T+1)+\rho f_T+\frac{\dot
f_R}{A^2}\left(\frac{2\dot{B}}{B}-\frac{\dot{A}}{A} \right)+
\frac{\ddot{f_R}}{A^2}-\frac{f-Rf_R}{2}\right.\\\label{15}&&\left.-\frac{f_R''}{B^2}
-\frac{f_R^\theta}{r^2B^2}\left(\frac{A^\theta}{A}-\frac{B^\theta}{B}
+\frac{C^\theta}{C}\right)-\frac{f_R'}{B^2}\left(\frac{A'}{A}-\frac{B'}{B}+\frac{C'}{C}\right)
\right], \\\nonumber G^{33}&=&\frac{1}{C^2f_R}\left[P_{zz}(f_T+1)+
\frac{\ddot{f_R}}{A^2}-\frac{ f_R^{\theta\theta}}{r^2B^2}+\rho
f_T-\frac{f-Rf_R}{2}-\frac{\dot f_R}{A^2} \left(\frac{\dot{A}}{A}
-\frac{\dot{2B}}{B}\right)\right.\\\label{16}&&\left.
-\frac{f_R''}{B^2}-\frac{f_R'}{B^2}\left(\frac{A'}{A}-\frac{2B'}{B}-\frac{1}{r}\right)
-\frac{f_R^\theta}{r^2B^2}\left(\frac{A^\theta}{A}-\frac{2B^\theta}{B}\right)\right].
\end{eqnarray}
The equation for Ricci scalar is found as
\begin{eqnarray}\nonumber
R&=&\frac{2}{A^2}\left[\frac{3\dot{A}}{A}\frac{\dot{B}}{B}-
3\left(\frac{\dot{B}}{B}\right)^2-\frac{2\ddot{B}}{B}
-\frac{\ddot{C}}{C}\right]\\\nonumber&&+\frac{2}{B^2}\left[\frac{A''}{A}+
\frac{A'C'}{AC}+\frac{B''}{B}-\frac{1}{r}\left(\frac{A'}{A}-\frac{B'}{B}-\frac{C'}{C}\right)-
\frac{B'^2}{B^2}\right.\\\label{17}&&\left.+\frac{C''}{C}+\frac{1}{r^2}
\left(\frac{A^{\theta\theta}}{A}+
\frac{B^{\theta\theta}}{B}+\frac{C^{\theta\theta}}{C}
-(\frac{{B^\theta}}{B})^2+\frac{A^\theta
C^\theta}{AC}\right)\right].
\end{eqnarray}

\subsection{Conservation Equations}

In $f(R, T)$ gravity, the divergence of energy momentum tensor leads
to the following expression
\begin{equation}\label{18}
\nabla^u T_{uv}=\frac{f_T (R, T)}{\kappa-f_T (R,
T)}\left([T_{uv}+\Theta_{uv}] \nabla^u\ln f_T (R,
T)+\nabla^u\Theta_{uv}-\frac{1}{2}g_{\mu\nu}\nabla^\mu T\right).
\end{equation}
It is clear from above equation that unlike GR the stress energy
tensor remains non-conserved in $f(R, T)$ gravity. This is because
of the direct matter geometry coupling, non-zero divergence induces
a force acting orthogonally to four velocity leading to the
non-geodesic path of test particles in non-minimally coupled
gravitational theories. Dynamics of a system can be explored by
considering conservation laws, divergence free gravitational
theories can be dealt by taking conservation of energy momentum
tensor. However, in our case we can not do so since $T_{uv}$ is
non-conserved that is why we take conservation of full form of field
tensor i.e., effective Einstein tensor.

Bianchi identities are used here to arrive at conservation equations
that are vital in establishment of evolution equation analytically.
Following equations are obtained from Bianchi identities
%\begin{eqnarray}\label{15}
%&&G^{uv}_{;v}V_{u}=0 \Rightarrow \left[\frac{1}{f_R}T^{0v}
%+\frac{1}{f_R}\overset{(D)}{T^{0v}}\right]_{;v}(-A)=0,
%\\\label{16}
%&&G^{uv}_{;v}K_{u}=0 \Rightarrow \left[\frac{1}{f_R}T^{1v}
%+\frac{1}{f_R}\overset{(D)}{T^{1v}}\right]_{;v}(B)=0,
%\\\label{17}
%&&G^{uv}_{;v}L_{u}=0 \Rightarrow \left[\frac{1}{f_R}T^{2v}
%+\frac{1}{f_R}\overset{(D)}{T^{2v}}\right]_{;v}(rB)=0,
%\end{eqnarray}
%on simplification, we have
\begin{eqnarray}\nonumber &&
G^{00}_{,0}+ G^{01}_{,1}+G^{02}_{,2}+G^{00}\left(\frac{2\dot{A}}{A}
+\frac{3\dot{B}}{B}\right)+G^{01}\left(\frac{3A'}{A}+\frac{2B'}{B}
+\frac{C'}{C}+\frac{1}{r}\right)\\\label{19}&&+G^{02}\left(\frac{3A^\theta}{A}+\frac{2B^\theta}{B}
+\frac{C^\theta}{C}\right)+G^{11}\frac{B\dot{B}}{A^2}+G^{22}\frac{r^2B\dot{B}}{A^2}
+G^{33}\frac{C\dot{C}}{A^2}=0,
\\\nonumber &&
G^{01}_{,0}+
G^{11}_{,1}+G^{12}_{,2}+G^{00}\frac{AA'}{B^2}+G^{01}\left(\frac{\dot{A}}{A}
+\frac{5\dot{B}}{B}\right)+G^{11}\left(\frac{A'}{A}+\frac{3B'}{B}
+\frac{C'}{C}\right.\\\label{20}&&\left.+\frac{1}{r}\right)+G^{12}\left(\frac{A^\theta}{A}
+\frac{4B^\theta}{B}
+\frac{C^\theta}{C}\right)-G^{22}\left(r+\frac{r^2B'}{B}\right)+G^{33}\frac{CC'}{B^2}=0,
\\\nonumber && G^{02}_{,0}+
G^{12}_{,1}+G^{22}_{,2}+G^{00}\frac{AA^\theta}{r^2B^2}+G^{02}\left(\frac{\dot{A}}{A}
+\frac{5\dot{B}}{B}\right)-\frac{B^\theta}{r^2B}G^{11}+\left(\frac{A'}{A}
\right.\\\label{21}&&\left.+\frac{4B'}{B}
+\frac{C'}{C}+\frac{3}{r}\right)G^{12}+G^{22}\left(\frac{A^\theta}{A}+\frac{3B^\theta}{B}
+\frac{C^\theta}{C}\right)-G^{33}\frac{CC^\theta}{r^2B^2}=0.
\end{eqnarray}
The notion $0, 1$ and $2$ stands for $t, r$ and $\theta$. One can
separate usual and dark source ingredients by substituting
components of Einstein tensor from Eqs.(\ref{10})-(\ref{16}) in
Eqs.(\ref{19})-(\ref{21}).

\section{Perturbation Scheme}

Theoretically and experimentally consistent $f(R, T)$ models can be
constructed, such models shall be compatible with local gravity
tests and cosmological constraints. These constraints are required
to meet for consistent solar system tests, matter domination phase
and stable high-curvature configuration. Generally, a consistent
gravitational model corresponds to the choice of parameters that are
in accordance with the observational scenarios \cite{49}. A viable
$f(R, T)$ ghost-free model shall have positive first and second
order derivatives for stability of cosmological perturbations
\cite{50}. In addition to this, any modified gravity model shall
agree with weak field limit and must be stable at semiclassical and
classical levels.

The viable models in $f(R, T )$ gravity have classification based on
fundamental form of $f$, as follows
\begin{itemize}
\item The model $f (R, T ) = R + 2f(T)$ describes
the cosmological model accompanying time dependent and effective
coupling, the term $2f(T)$ characterizes the interaction of matter
and curvature interaction in the extended gravitational action.
\item $f(R, T ) = f_1 + f_2$, $f_1$ and $f_2$
are arbitrary functions of $R$ and $T$, respectively appearing
explicitly.
\item $f(R, T ) = f_1(R) + f_2(R)f_3(T)$ denotes the most general
class of $f(R,T)$ models, carrying implicit form of $R$ and $T$ in
extended gravitational action.
\end{itemize}

Herein, we are dealing with axial system analytically, in order to
arrive at some fruitful conclusion avoiding complexities we select
$f(R, T ) = f_1(R) + f_2(T)$ type. More particularly, we take
$f(R,T)=R+\alpha R^2+\lambda T$, where $\alpha$ and $\lambda$ denote
positive constants. The origin of such constraints on model comes
from the argument that non-linear terms of $T$ in the model
complicates the field equations that can not be tackled by analytic
approach. The $f(R, T)$ models containing non-linear enteries of $T$
can be handled by numerical simulations but such analysis provide
outcomes for a specific model. Whereas, more generic results can be
found for a variety of models with arbitrary values of $\alpha$ and
$\lambda$ by using analytic approach. .

Following scheme for perturbation of physical variables is employed
to monitor the fluctuations in the gravitating object with the time
transition. Metric coefficients and Ricci scalar are perturbed by
using following pattern
\begin{eqnarray}\label{22'}
L(t,r,\theta)&=&L_0(r,\theta)+\epsilon D(t)l(r,\theta).
\end{eqnarray}
While energy density and pressures stresses are perturbed as follows
\begin{eqnarray}\label{22''}
\rho(t,r,\theta)&=&\rho_0(r,\theta)+\epsilon
{\bar{\rho}(t,r,\theta)},
\\
\label{26'} P_{ij}(t,r,\theta)&=&P_{ij0}(r,\theta)+\epsilon
{\bar{P}_{ij}(t,r,\theta)}.
\end{eqnarray}
Variation in $f(R,T)$ model is taken in the following form
\begin{eqnarray}\nonumber
f(R, T)&=&[R_0(r,\theta)+\alpha R_0^2(r,\theta)+\lambda
T_0(r,\theta)]+\epsilon D(t)e(r,\theta)[1\\\label{51'}&+&2\alpha
R_0(r,\theta)],\\\label{52'} f_R&=&1+2\alpha R_0(r,\theta)+\epsilon
2\alpha D(t)e(r,\theta),\\\label{52''} f_T&=&\lambda,
\end{eqnarray}
where $0<\epsilon\ll1$. Application of the first order perturbation on dynamical equations (\ref{19})-(\ref{21})
implies
\begin{eqnarray}\nonumber&&
\left[\dot{\bar{\rho}} +\left\{\rho_0\left(\frac{a}{A_0}+\frac{3\lambda_1b}{B_0}\right)+
\frac{\lambda_1b}{B_0}(P_{xx0}+P_{yy0}+P_{zz0})+Z_{1p}\right\}\dot{D}\right]=0,
\\\label{33}&&
\\\nonumber&&
\left[\lambda_1\bar{P_{xx}}+\lambda\bar{\rho}-2(\lambda_1{P_{xx0}}+\lambda\rho_0)\left(\frac{b}{B_0}
+\frac{e\alpha}{I}\right)D\right]_{,1}+\left(\lambda_1\bar{P_{xx}}+\lambda\bar{\rho}\right)
\left(\frac{A_0'}{A_0}+\frac{3B_0'}{B_0}
\right.\\\nonumber&&\left.+\frac{C_0'}{C_0}
+\frac{1}{r}\right)+\frac{1}{r^2}\left[\lambda_1\bar{P_{xy}}-2\left(\frac{2b}{B_0}
+\frac{e\alpha}{I}\right){P_{xy0}}D\right]_{,2}+\frac{\lambda_1\bar{P_{xy}}}{r^2B_0^2}\left(\frac{A_0^\theta}{A_0}
+4\frac{B_0^\theta}{B_0}+\frac{C_0^\theta}{C_0}\right)+\\\nonumber&&\left(\lambda_1\bar{P_{yy}}+\lambda\bar{\rho}\right)
\left(\frac{1}{r}+\frac{B_0'}{B_0}\right)+\left(\lambda_1\bar{P_{zz}}+\lambda\bar{\rho}\right)\frac{C_0'}{C_0}+
D\left[(\lambda_1{P_{xx0}}+\lambda\rho_0)\left(\left(\frac{a}{A_0}\right)'+
\right.\right. \\\nonumber&&\left.\left.+4\left(\frac{b}{B_0}\right)'-\left(\frac{2b}{B_0}+\frac{e\alpha}{I}\right)
\left(\frac{A_0'}{A_0}+\frac{3B_0'}{B_0}+\frac{C_0'}{C_0}+\frac{1}{r}\right)\right)+(\lambda_1{P_{yy0}}
+\lambda\rho_0)
\left(\left(\frac{b}{B_0}\right)'\right. \right. \\\nonumber&&\left.\left.-\left(\frac{2b}{B_0}+\frac{e\alpha}{I}\right)\frac{B_0'}{B_0}
\right)\left(\frac{1}{r}+\frac{B_0'}{B_0}\right)+(\lambda_1{P_{zz0}}+\lambda\rho_0)
\left(\left(\frac{c}{C_0}\right)'-\left(\frac{2b}{B_0}+\frac{e\alpha}{I}\right)\frac{C_0'}{C_0}
\right)\right. \\\label{34}&&\left.+\lambda_1{P_{xy0}}\left(\left(\frac{a}{A_0}\right)^\theta+
5\left(\frac{b}{B_0}\right)^\theta-\left(\frac{2b}{B_0}
+\frac{e\alpha}{I}\right)\frac{C_0^\theta}{C_0}\right)\right]+Z_{2p}=0,
\end{eqnarray}
\begin{eqnarray}
\nonumber&&
\left[\lambda_1\bar{P_{yy}}+\lambda\bar{\rho}-2(\lambda_1{P_{yy0}}+\lambda\rho_0)\left(\frac{b}{B_0}
+\frac{e\alpha}{I}\right)D\right]_{,2}+\left[\frac{1}{r^2B_0^4I}\lambda_1\bar{P_{xy}}\right]'
+\bar{\rho}\frac{A_0^\theta}{A_0}\\\nonumber&&
+\left(\lambda_1\bar{P_{xx}}+\lambda\bar{\rho}\right)\frac{B_0^\theta}{B_0}+
\lambda_1\bar{P_{xy}}
\left(\frac{A_0'}{A_0}+\frac{4B_0'}{B_0}+\frac{C_0'}{C_0}
+\frac{3}{r}\right)+\left(\lambda_1\bar{P_{yy}}+\lambda\bar{\rho}\right)\left(\frac{A_0^\theta}{A_0}
\right.\\\nonumber&&\left.
+3\frac{B_0^\theta}{B_0}+\frac{C_0^\theta}{C_0}\right)+\left(\lambda_1\bar{P_{zz}}
+\lambda\bar{\rho}\right)\frac{C_0^\theta}{C_0}
+
D\left[\rho_0\left(\left(\frac{a}{A_0}\right)^\theta-2\left(\frac{b}{B_0}
+\frac{e\alpha}{I}\right)\frac{A_0^\theta}{A_0}\right)\right.\\\nonumber&&\left.+
\lambda_1{P_{xy0}}\left(\left(\frac{a}{A_0}\right)'+
+5\left(\frac{b}{B_0}\right)'-\left(\frac{4b}{B_0}+\frac{2e\alpha}{I}\right)
\left(\frac{A_0'}{A_0}+\frac{4B_0'}{B_0}+\frac{C_0'}{C_0}
\right.\right.\right.\\\nonumber&&\left.\left.\left.+\frac{3}{r}\right)\right)+(\lambda_1{P_{xx0}}
+\lambda\rho_0)\left(
4\left(\frac{b}{B_0}\right)^\theta-\left(\frac{2b}{B_0}
+\frac{e\alpha}{I}\right)\frac{B_0^\theta}{B_0}\right)+(\lambda_1{P_{yy0}}
+\lambda\rho_0)\right.\\\nonumber&&\left.\times
\left(\left(\frac{a}{A_0}\right)^\theta+
4\left(\frac{b}{B_0}\right)^\theta-\left(\frac{2b}{B_0}
+\frac{e\alpha}{I}\right)\left(\frac{A_0^\theta}{A_0}+3\frac{B_0^\theta}{B_0}
+\frac{C_0^\theta}{C_0}\right)\right)
\right.\\\label{35}&&\left.+(\lambda_1{P_{zz0}}
+\lambda\rho_0)\left(\left(\frac{b}{B_0}\right)^\theta-\left(\frac{2b}{B_0}
+\frac{e\alpha}{I}\right)\frac{C_0^\theta}{C_0}\right)\right]+Z_{3p}=0,
\end{eqnarray}
where expressions for $Z_{1p}, Z_{2p}$ and $Z_{3p}$ given in
appendix. We take, $I = 1+2\alpha R_0$ and $J = e2\alpha R_0$ for
the sake of simplicity. The value of energy density $\bar{\rho}$ is
derived from Eq.(\ref{33}) which turns out to be
\begin{eqnarray}\nonumber&&
\bar{\rho}=-\left\{\rho_0\left(\frac{a}{A_0}+\frac{3\lambda_1b}{B_0}\right)+
\frac{\lambda_1b}{B_0}(P_{xx0}+P_{yy0}+P_{zz0})+Z_{1p}\right\}D.\\\label{36}&&
\end{eqnarray}

The expression for $\rho$ and pressure stresses can be related as
\cite{1}
\begin{equation}\label{39}
\bar{P}_i=\Gamma\frac{p_{i0}}{\rho_0+p_{i0}}\bar{\rho}.
\end{equation}
where $\Gamma$ represents the variation of anisotropic pressure
stresses with the varying energy density and index $i=xx, yy, xy,
zz$. Making use of Eq.(\ref{39}) in Eq.(\ref{36}) and some algebraic
manipulations provide linearly perturbed anisotropic stresses. An
ordinary differential equation is obtained by using linearly
perturbed form of Ricci scalar having solution as follows
\begin{equation}\label{38}
D(t)=-e^{\sqrt{Z_4}t}.
\end{equation}
The expression for $Z_4$ is given in appendix, Eq.(\ref{38}) is
holds for positive values of $Z_4$.

\section{Weak Field Limit}

In this section, we found the entries belonging to Newtonian (N) and
post Newtonian (pN) eras. Equations (\ref{38}) and (\ref{39}) are
inserted in Eq. (\ref{34}) to arrive at evolution equation from
which instability criterion is developed in terms of adiabatic
index.

\subsection{N Approximation}

In this regime, we take $A_0=1,~B_0=1$, $\rho_0\gg p_{i0}$ and
Schwarzschild coordinates $C_0=r$ in evolution equation, whose
outcome reveals the following relation
\begin{equation}\label{n}
\Gamma <
\frac{\lambda N_0'-\frac{3}{r}N_0-2\lambda(\rho_0N_2)'-\frac{2}{r}(P_{xy0}N_2)^\theta+\lambda N_2 N_3-\frac{2}{r}N_2+\lambda P_{xy0}N_4+Z_{2^{N}_{p}}}{\lambda_1(P_{xx0}N_1)'+\frac{\lambda_1}{r^2}
(P_{xy0}N_1)^\theta-\frac{1}{r}N_1(P_{xx0}+P_{yy0}+P_{zz0})},
\end{equation}
where $Z_{2^{N}_{p}}$ represent those terms of $Z_{2p}$ that belongs
to the N-limit. Furthermore
\begin{eqnarray}\nonumber &&
N_0=-\left\{\rho_0N_1+
\frac{\lambda_1c}{r}(P_{xx0}+P_{yy0}+P_{zz0})+Z_{1^{N}_{p}}\right\},
\\\nonumber&&
N_1=a+\frac{3\lambda_1c}{r}, \quad N_2=\frac{c}{r}+\frac{\alpha e}{I}
\\\nonumber&&
N_3=a'+4b'+2(\frac{c}{r})', \quad N_4=a^\theta+4b^\theta+\frac{c^\theta}{r}.
\end{eqnarray}
Both usual matter and dark source entries are taking part in above
inequality for $\Gamma$, gravitational sources maintain stability
for those values of physical parameters for which the inequality
(\ref{n}) is satisfied. All entries appearing in (\ref{n}) shall
remain positive, to fulfil this requirement we need to constrain
some of the physical parameters. The restrictions in N-limit are
\begin{eqnarray}\nonumber &&
P_{xx0}+P_{yy0}+P_{zz0}<\frac{\lambda_1r}{N_1}((P_{xx0}N_1)'+\frac{1}{r^2},
\quad (P_{xy0}N_2)^\theta<-2\lambda (\rho_0N_2)'.
\end{eqnarray}
Stability of the gravitating system distorts whenever above
constraints are violated that leads to the instability range of
collapsing stars.

\subsection{pN Approximation}

In pN limit, we take $A_0=1-\frac{m_0}{r}$ and
$B_0=1+\frac{m_0}{r}$, use of these assumptions in evolution
equation leads to following inequality for $\Gamma$
\begin{equation}\label{pn}
\Gamma < \frac{\lambda
N_{10}'+N_9N_{10}-2\lambda(\rho_0N_6)'-\frac{2}{r}(P_{xy0}N_6)^\theta+\lambda
\rho_0 N_7-\frac{3}{r}N_6+\lambda
P_{xy0}N_8+Z_{2^{pN}_{p}}}{\lambda_1(P_{xx0}N_5)'+\frac{\lambda_1}{r^2}
(P_{xy0}N_5)^\theta-\frac{1}{r}N_5(P_{xx0}+P_{yy0}+P_{zz0})+N_{11}},
\end{equation}
where
\begin{eqnarray}\nonumber &&
N_5=\left(\frac{ar}{r-m_0}+\frac{2\lambda_1br}{r+m_0}+\frac{\lambda_1c}{r}\right), \quad N_6=\frac{2\lambda_1br}{r+m_0}+\frac{e\alpha}{I},
\\\nonumber&&
N_7=\left(\frac{ar}{r-m_0}\right)'+4\left(\frac{br}{r+m_0}\right)'
+\left(\frac{2c}{r}\right)'-N_6\left(\frac{2}{r}+\left(\frac{m_0}{r}\right)'\frac{3r}{r+m_0}\right),
\\\nonumber&&
N_8=\left(\frac{ar}{r-m_0}\right)^\theta+\left(\frac{br}{r+m_0}\right)^\theta+\left(\frac{c}{r}\right)^\theta,
\quad
N_9=\left(\frac{3}{r}+\left(\frac{m_0}{r}\right)'\frac{3r}{r+m_0}\right),
\\\nonumber&&
N_{10}=-\left\{\rho_0N_5+
\frac{2\lambda_1br}{r+m_0}(P_{xx0}+P_{yy0})+\frac{\lambda_1c}{r}P_{zz0}+Z_{1^{pN}_{p}}\right\},
\\\nonumber&&
N_{11}=\frac{P_{xy0}N_5}{(r+m_0)^2}\left(\left(\frac{ar}{r-m_0}\right)^\theta
+\left(\frac{4br}{r+m_0}\right)^\theta\right).
\end{eqnarray}

Likewise N-limit the restrictions in metric functions and dark
source entries are required to achieve stability of gravitating
sources. System is exposed to gravitational collapse whenever the
inequality (\ref{pn}) breaks down. Some of the constraints are
\begin{eqnarray}\nonumber &&
P_{xy}\left[\frac{2\lambda_1br}{r+m_0}+\frac{e\alpha}{I}\right]<0,
\\\nonumber && \left\{\rho_0N_5+
\frac{2\lambda_1br}{r+m_0}(P_{xx0}+P_{yy0})+\frac{\lambda_1c}{r}P_{zz0}+Z_{1^{pN}_{p}}\right\}<0.
\end{eqnarray}

\section{Summary}

Study of non-spherical symmetries provide deep insight of more
realistic settings such as weak lensing, large scale structures,
Planck data, cosmic microwave background etc. In this work, we have
studied the impact of restricted axial symmetry (ignoring reflection
and rotation) on anisotropic $f(R,T$ gravity model evolving under
shear-free condition. Restricted character of spacetime incorporates
three independent metric functions depending on $t$, $r$ and
$\theta$, that leads to the vorticity-free case. We choose, $f(R,
T)=R+\alpha R^2+\lambda T$ with positive values of $\alpha$ and
$\lambda$ as a viable $f(R, T)$ model to discuss the instability
range.

The variation of extended gravitational action (\ref{1}) is
considered to obtain the modified field equations in $f(R, T)$
gravity. We make use of shear-free condition to evaluate components
of modified field equations. The energy momentum tensor remains
non-conserved in $f(R, T)$ gravity that is why we take conservation
of modified Einstein tensor. The conservation equations lead to
highly non-linear complicated equations. To tackle with the
complexities of dynamical equations, we implement linear
perturbation on all physical parameters. Application of linear
perturbation yield equations for perturbed energy density that is
further used to evaluate perturbed anisotropic stresses.

Insertion of perturbed stresses and energy density in second Bianchi
identity leads to the collapse equation carrying adiabatic index
that elaborates the variation of anisotropic pressures with energy
density. It is found that imposed shear-free condition reduces the
entries with negative sign in collapse equation that leads to less
restricted and enhanced regions of stability. Thus it can be
remarked that vanishing shear somehow delays the collapsing
phenomenon. Moreover, weak field limit is checked by evaluating
expression for adiabatic index in Newtonian and post Newtonian
approximations. Corrections to GR can be recovered by taking
$\alpha\rightarrow0, \lambda\rightarrow0$ in evolution equation.
While $\lambda\rightarrow0$ reduces results in $f(R)$ gravity for
Starobinsky model.

\section*{Appendix}

\begin{eqnarray}\nonumber&&
Z_{1p}=\frac{e}{2}-A_0^2\left\{\frac{1}{A_0^2B_0^2I^2}\left(
(2\alpha e R_0)'(1-\frac{b}{B_0})-2\alpha eR_0\frac{A_0'}{A_0}\right)\right\}_{,1}-\frac{A_0^2}{r^2}\left\{\frac{2}{A_0^2B_0^2I^2}\right.\\\nonumber&&\left.\times\left(
(\alpha e R_0)^\theta(1-\frac{b}{B_0})-(\alpha e R_0)\frac{A_0^\theta}{A_0}\right)\right\}_{,2}
+\frac{\alpha^2R_0^3}{I}+\frac{1}{B_0^2}
\left[\frac{(e^\theta (2\alpha e R_0))^\theta}{r^2}-4\alpha\right.\\\nonumber&&\left.\times\left((R_0R_0')'+\frac{(R_0R_0^\theta)^\theta}{r^2}\right)
\left(\frac{a}
{A_0}+\frac{b}{B_0}+\frac{\alpha e R_0}{I}\right)
+I'\left\{-\left(\frac{b}{B_0}\right)'\right.\right.\\\nonumber&&\left.\left.
-\frac{b}{B_0}\left(\frac{3A_0'}{A_0}+\frac{2B_0'}{B_0}-\frac{C_0'}{C_0}-\frac{4}{r}\right)
-\frac{c}{C_0}\left(\frac{A_0'}{A_0}-\frac{C_0'}{C_0}-\frac{1}{r}\right)+
\frac{(2\alpha e R_0)}{I}\left(\frac{C_0'}{C_0}+\frac{2B_0'}{B_0}\right.\right.\right.\\\nonumber&&\left.\left.\left.
-\frac{3}{r}\right)\right\}
+(e'(2\alpha e R_0))'
+\frac{I^\theta}{r^2}\left\{
-\left(\frac{b}{B_0}\right)^\theta-\frac{b}{B_0}\left(\frac{3A_0^\theta}{A_0}+\frac{2B_0^\theta}{B_0}
-\frac{C_0^\theta}{C_0}\right)
\right.\right.\\\nonumber&&\left.\left.
+
\frac{(2\alpha e R_0)}{I}\left(\frac{C_0^\theta}{C_0}+\frac{2B_0^\theta}{B_0}\right)\right\}+(2\alpha e R_0)'\left(\frac{C_0'}{C_0}-\frac{2B_0'}{B_0}+\frac{1}{r}\right)
\right.\\\nonumber&&\left.
+\frac{(2\alpha e R_0)^\theta}{r^2}\left(\frac{C_0^\theta}{C_0}
-\frac{2B_0^\theta}{B_0}\right)+\left(\frac{2a}{A_0}+\frac{b}{B_0}\right)\left(I''
+\frac{I^{\theta\theta}}{r^2}\right)
-\left(\frac{3A_0'}{A_0}+\frac{2B_0'}{B_0}+\frac{1}{r}\right.\right.\\\nonumber&&\left.\left.
+\frac{C_0'}{C_0}
\right)\left(\frac{(2\alpha e R_0)'}{I}(1-\frac{b}{B_0})
-\frac{A_0'}{A_0}\frac{(2\alpha e R_0)}{I}\right)
+\left(\frac{A_0^\theta}{A_0}
\frac{(2\alpha e R_0)}{I}-\frac{(2\alpha e R_0)^\theta}{I}(1\right.\right.\\\label{a1}&&\left.\left.
-\frac{b}{B_0})\right)\left(\frac{3A_0^\theta}{A_0}+\frac{C_0^\theta}{C_0}
+\frac{2B_0^\theta}{B_0}\right)\right],
\end{eqnarray}
\begin{eqnarray}\nonumber&&
Z_{2p}=\left[\left[\frac{1}{IB_0^2}\left\{\frac{\ddot{D}}{DA^2_0}-\frac{1}{B_0^2}
\left\{\frac{eB_0^2}{2}+I'\left(\frac{a}{A_0}\right)'
\right.\right.\right.\right.\\\nonumber&&\left.\left.\left.\left.
+\left(J'-\frac{2b}{B_0}I'\right)\left(\frac{A_0'}{A_0}+\frac{C_0'}{C_0}
-\frac{B_0'}{B_0}-\frac{1}{r}\right)+\frac{1}{r^2}\left(J^{\theta\theta}+\left(J^\theta
-\frac{2b}{B_0}I^\theta\right)
\left(\frac{A_0^\theta}{A_0}
\right.\right.\right.\right.\right.\right.\\\nonumber&&\left.\left.\left.\left.\left.\left.-
\frac{3B_0^\theta}{B_0}
+\frac{C_0^\theta}{C_0}\right)+
\frac{2b}{B_0}I^{\theta\theta}+I^\theta
\left(\left(\frac{a}{A_0}\right)^\theta-2\left(\frac{b}{B_0}\right)^\theta
\right)\right)\right\}\right\}\right]_{,1}\right.\\\nonumber&&\left.+
\left[\frac{1}{r^2IB_0^4}\left\{J'^{\theta}+\left(\frac{b}{B_0}\right)^\theta I'+
J^{\theta}\left(\frac{B_0'}{B_0}+\frac{1}{r}\right)
-\left(\frac{b}{B_0}\right)'I^\theta\right\}\right]_{,2}\right]IB_0^4
\\\nonumber&&-e\frac{B_0'}{B_0}+\frac{A_0'}{A_0}\left[J''+\frac{J^{\theta\theta}}{r^2}
-\frac{2b}{B_0}\left(I''+\frac{I^{\theta\theta}}{r^2}\right)+\left(J'
-\frac{2b}{B_0}I'\right)\left(\frac{C_0'}{C_0}
-\frac{2B_0'}{B_0}\right.\right.\\\nonumber&&\left.\left.
+\frac{1}{r}\right)-\frac{1}{r^2}\left\{I^\theta
\left(\frac{b}{B_0}\right)^\theta
-\left(J^\theta\right.\right.\right.\\\nonumber&&\left.\left.\left.
-\frac{2b}{B_0}I^\theta\right)\left(\frac{C_0^\theta}{C_0}-\frac{2B_0^\theta}{B_0}\right)\right\}\right]+
\left(\frac{(aA_0)'}{A_0^2}-\frac{2b}{B_0}\frac{A_0'}{A_0}\right)\left(\frac{\alpha R_0^2B_0^2}{2}+I''+
I'\left(\frac{C_0'}{C_0}\right.\right.\\\nonumber&&\left.\left.
-\frac{2B_0'}{B_0}+\frac{1}{r}\right)+\frac{I^{\theta\theta}}{r^2}
+\frac{I^{\theta}}{r^2}\left(\frac{C_0^\theta}{C_0}-\frac{2B_0^\theta}{B_0}\right)\right)-
\left\{\frac{\alpha R_0^2B_0^2}{2}+I'\left(\frac{A_0'}{A_0}+\frac{C_0'}{C_0}
-\frac{B_0'}{B_0}\right.\right.\\\nonumber&&\left.\left.-\frac{1}{r}\right)+\frac{I^{\theta\theta}}{r^2}
+\frac{I^{\theta}}{r^2}\left(\frac{A_0^\theta}{A_0}+\frac{C_0^\theta}{C_0}-\frac{3B_0^\theta}{B_0}\right)\right\}
\left(\left(\frac{a}{A_0}\right)'+4\left(\frac{b}{B_0}\right)'
\right)\\\nonumber&&-\left(\frac{A_0'}{A_0}+\frac{C_0'}{C_0}
+\frac{3B_0'}{B_0}+\frac{1}{r}\right)\left\{I'\left(\frac{a}{A_0}\right)'
+\left(\frac{A_0'}{A_0}
-\frac{B_0'}{B_0}\right.\right.\\\nonumber&&\left.\left.
+\frac{C_0'}{C_0}-\frac{1}{r}\right)\left(J'-\frac{2b}{B_0}I'\right)+\frac{1}{r^2}\left(J^{\theta\theta}+\left(J^\theta
-\frac{2b}{B_0}I^\theta\right)
\left(\frac{A_0^\theta}{A_0}-
\frac{3B_0^\theta}{B_0}
+\frac{C_0^\theta}{C_0}\right)\right.\right.\\\nonumber&&\left.\left.+
\frac{2b}{B_0}I^{\theta\theta}+I^\theta
\left(\left(\frac{a}{A_0}\right)^\theta
-2\left(\frac{b}{B_0}\right)^\theta
\right)\right)\right\}
-\left[\left(\left(\frac{a}{A_0}\right)^\theta
\right.\right.\\\nonumber&&\left.\left.+5\left(\frac{b}{B_0}\right)^\theta
\right)\left(I'^\theta +\frac{B_0^\theta}{B_0}I'+I^\theta\left(\frac{B_0'}{B_0}
+\frac{1}{r}\right)\right)-\left(\frac{A_0^\theta}{A_0}+
\frac{4B_0^\theta}{B_0}
+\frac{C_0^\theta}{C_0}\right)\left(\frac{B_0^\theta}{B_0}J'
\right.\right.\\\nonumber&&\left.\left.
-J'^\theta- I'\left(\frac{b}{B_0}\right)^\theta -J^\theta\left(\frac{B_0'}{B_0}
+\frac{1}{r}\right)+I^\theta\left(\frac{b}{B_0}\right)'\right)\right]\frac{1}{r^2}-
\left(\frac{B_0'}{B_0}+\frac{1}{r}\right)\left[\frac{B_0^2}{A_0^2}\frac{\ddot{D}}{D}J
\right.\\\nonumber&&\left.-J''+\frac{2b}{B_0}I''-I'\left(\frac{a}{A_0}\right)'
-\left(J'-\frac{2b}{B_0}I'\right)\left(\frac{A_0'}{A_0}+\frac{C_0'}{C_0}\right.\right.
\end{eqnarray}
\begin{eqnarray}
\nonumber&&\left.\left.-\frac{B_0'}{B_0}\right)+\frac{1}{r^2}\left(I^\theta
\left(\frac{a}{A_0}\right)^\theta-\left(J^\theta -\frac{2b}{B_0}I^\theta\right)
\left(\frac{A_0^\theta}{A_0}-\frac{B_0^\theta}{B_0}
\right.\right.\right.\\\nonumber&&\left.\left.\left.+\frac{C_0^\theta}{C_0}\right)\right)\right]
+\left(\frac{b}{B_0}\right)'\left[\frac{LB_0^2}{2}+I'
\left(\frac{A_0'}{A_0}+\frac{C_0'}{C_0}-\frac{B_0'}{B_0}\right)-\frac{I^\theta}{r^2}
\left(\frac{A_0^\theta}{A_0}-\frac{B_0^\theta}{B_0}
+\frac{C_0^\theta}{C_0}\right)\right.\\\nonumber&&\left.-I''\right]+
\frac{C_0'}{C_0}\left[J''-\frac{2b}{B_0}I''+I'\left(\left(\frac{a}{A_0}\right)'-\left(\frac{2b}{B_0}\right)'\right)
+\left(J'-\frac{2b}{B_0}I'\right)\left(\frac{A_0'}{A_0}-\frac{B_0'}{B_0}\right.\right.\\\nonumber&&\left.\left.
+\frac{1}{r}\right) +\frac{1}{r^2}\left\{J^{\theta\theta}+
\left(J^\theta -\frac{2b}{B_0}I^\theta\right)
\left(\frac{A_0^\theta}{A_0}-\frac{2B_0^\theta}{B_0}\right)+I^\theta
\left(\left(\frac{a}{A_0}\right)^\theta-2\left(\frac{b}{B_0}\right)^\theta
\right)\right.\right.\\\nonumber&&\left.\left.
-\frac{2b}{B_0}I^{\theta\theta}\right\}\right]+\left(\frac{(cC_0)'}{C_0^2}-\frac{2b}{B_0}\frac{C_0'}{C_0}\right)
\left[I'
\left(\frac{A_0'}{A_0}-\frac{2B_0'}{B_0}+\frac{1}{r}\right)-\frac{I^\theta}{r^2}
\left(\frac{A_0^\theta}{A_0}-\frac{2B_0^\theta}{B_0}\right)
\right.\\\label{a2}&&\left.+\frac{\alpha R_0^2B_0^2}{2}+I''+\frac{I^{\theta\theta}}{r^2}\right]-
\frac{\ddot{D}B_0^2}{DA^2_0I}\left(J'-\frac{A_0'}{A_0}J-\frac{b}{B_0}I'\right),
\\\nonumber&&
Z_{3p}=Ir^2B_0^4\left[\left[\frac{1}{r^2IB_0^4}\left\{J'^{\theta}+\left(\frac{b}{B_0}\right)^\theta
I'+ J^{\theta}\left(\frac{B_0'}{B_0}+\frac{1}{r}\right)
-\left(\frac{b}{B_0}\right)'I^\theta\right\}\right]_{,1}\right.\\\nonumber&&\left.+
\frac{\ddot{D}B_0^2}{DA^2_0I}\left(\frac{A_0^\theta}{A_0}J+\frac{b}{B_0}I^\theta-J^\theta\right)
+\left[\frac{1}{Ir^2B_0^4}\left\{\frac{\ddot{D}B_0^2}{DA_0^2}J
-J''+\frac{2b}{B_0}I''+\left(\frac{2b}{B_0}I'\right.\right.\right.\right.
\\\nonumber&&\left.\left.\left.\left.-J'\right)\left(\frac{A_0'}{A_0}+\frac{C_0'}{C_0}
-\frac{B_0'}{B_0}\right)-I'\left(\frac{a}{A_0}\right)'
+\frac{1}{r^2}\left(\left(\frac{2b}{B_0}I^\theta
\right.\right.\right.\right.\right.
\\\nonumber&&\left.\left.\left.\left.\left.-J^\theta
\right)
\left(\frac{A_0^\theta}{A_0}-\frac{B_0^\theta}{B_0}+\frac{C_0^\theta}{C_0}\right)-I^\theta
\left(\frac{a}{A_0}\right)^\theta\right)\right\}\right]_{,2}\right]
\\\nonumber&&-e\frac{B_0^\theta}{B_0}+\frac{A_0^\theta}{A_0}\left[J''+\frac{J^{\theta\theta}}{r^2}
-\frac{2b}{B_0}\left(I''+\frac{I^{\theta\theta}}{r^2}\right)+\left(J'
-\frac{2b}{B_0}I'\right)\left(\frac{C_0'}{C_0}
-\frac{2B_0'}{B_0}\right.\right.\\\nonumber&&\left.\left.
+\frac{1}{r}\right)-\frac{1}{r^2}\left\{I^\theta
\left(\frac{b}{B_0}\right)^\theta
-\left(J^\theta\right.\right.\right.\\\nonumber&&\left.\left.\left.
-\frac{2b}{B_0}I^\theta\right)\left(\frac{C_0^\theta}{C_0}-\frac{2B_0^\theta}{B_0}\right)\right\}\right]
+\left(\frac{(aA_0)^\theta}{A_0^2}-\frac{2b}{B_0}\frac{A_0^\theta}{A_0}\right)\left(\frac{\alpha R_0^2B_0^2}{2}+I''+
I'\left(\frac{C_0'}{C_0}\right.\right.
\end{eqnarray}
\begin{eqnarray}\nonumber&&\left.\left.
-\frac{2B_0'}{B_0}+\frac{1}{r}\right)+\frac{I^{\theta\theta}}{r^2}
+\frac{I^{\theta}}{r^2}\left(\frac{C_0^\theta}{C_0}-\frac{2B_0^\theta}{B_0}\right)\right)
-\left(\frac{b}{B_0}\right)^\theta\left\{\frac{\alpha R_0^2B_0^2}{2}+I'\left(\frac{A_0'}{A_0}+\frac{C_0'}{C_0}
\right.\right.\\\nonumber&&\left.\left.-\frac{B_0'}{B_0}+\frac{1}{r}\right)+\frac{I^{\theta\theta}}{r^2}
+\frac{I^{\theta}}{r^2}\left(\frac{A_0^\theta}{A_0}+\frac{C_0^\theta}{C_0}-\frac{3B_0^\theta}{B_0}\right)\right\}
-\frac{B_0^\theta}{B_0}\left\{I'\left(\frac{a}{A_0}\right)'
\right.\\\nonumber&&\left.
 +\left(\frac{A_0'}{A_0}
-\frac{B_0'}{B_0}
+\frac{C_0'}{C_0}-\frac{1}{r}\right)\left(J'-\frac{2b}{B_0}I'\right)+\frac{1}{r^2}\left(J^{\theta\theta}
-\left(\frac{2b}{B_0}I^\theta-\right.\right.\right.\\\nonumber&&\left.\left.\left.J^\theta\right)
\left(\frac{A_0^\theta}{A_0}- \frac{3B_0^\theta}{B_0}
+\frac{C_0^\theta}{C_0}\right)+
\frac{2b}{B_0}I^{\theta\theta}+I^\theta
\left(\left(\frac{a}{A_0}\right)^\theta
-2\left(\frac{b}{B_0}\right)^\theta \right)\right)\right\}
\\\nonumber&&
-\frac{1}{r^2}\left[\left(\left(\frac{a}{A_0}\right)'
+2\left(\frac{b}{B_0}\right)'
\right)\left(I'^\theta +\frac{B_0^\theta}{B_0}I'+I^\theta\left(\frac{B_0'}{B_0}
+\frac{1}{r}\right)\right)-\left(\frac{A_0'}{A_0}\right.\right.\\\nonumber&&\left.\left.+
\frac{4B_0'}{B_0}
+\frac{C_0'}{C_0}\right)\left(\frac{B_0^\theta}{B_0}J'
-J'^\theta- I'\left(\frac{b}{B_0}\right)^\theta -J^\theta\left(\frac{B_0'}{B_0}
+\frac{1}{r}\right)+I^\theta\left(\frac{b}{B_0}\right)'\right)\right]
\\\nonumber&&+\left(\frac{A_0^\theta}{A_0}+
\frac{3B_0^\theta}{B_0}
+\frac{C_0^\theta}{C_0}\right)\left[\frac{B_0^2}{A_0^2}\frac{\ddot{D}}{D}J+
\frac{2b}{B_0}I''-I'\left(\frac{a}{A_0}\right)'
\right.\\\nonumber&&\left.-J''
-\left(J'-\frac{2b}{B_0}I'\right)\left(\frac{A_0'}{A_0}+\frac{C_0'}{C_0}
-\frac{B_0'}{B_0}\right)-\frac{1}{r^2}\left(\left(J^\theta
-\frac{2b}{B_0}I^\theta\right)
\left(\frac{A_0^\theta}{A_0}-\frac{B_0^\theta}{B_0}
\right.\right.\right.
\\
\nonumber&&\left.\left.\left.+\frac{C_0^\theta}{C_0}\right)-I^\theta
\left(\frac{a}{A_0}\right)^\theta\right)\right]+
\left(\left(\frac{a}{A_0}\right)^\theta
\right.\\\nonumber&&\left.+4\left(\frac{b}{B_0}\right)^\theta
\right)\left[\frac{\alpha R_0^2B_0^2}{2}+I'
\left(\frac{A_0'}{A_0}+\frac{C_0'}{C_0}-\frac{B_0'}{B_0}\right)-\frac{I^\theta}{r^2}
\left(\frac{A_0^\theta}{A_0}-\frac{B_0^\theta}{B_0}
+\frac{C_0^\theta}{C_0}\right)-I''\right]
\\\nonumber&&-\frac{C_0^\theta}{C_0}\left[J''-\frac{2b}{B_0}I''
+I'\left(\left(\frac{a}{A_0}\right)'-\left(\frac{2b}{B_0}\right)'\right)
+\left(J'-\frac{2b}{B_0}I'\right)\left(\frac{A_0'}{A_0}-\frac{B_0'}{B_0}\right.\right.\\\nonumber&&\left.\left.
+\frac{1}{r}\right)+\frac{1}{r^2}\left\{\left(J^\theta-\frac{2b}{B_0}I^\theta\right)
\left(\frac{A_0^\theta}{A_0}-\frac{2B_0^\theta}{B_0}\right)+I^\theta
\left(\left(\frac{a}{A_0}\right)^\theta-2\left(\frac{b}{B_0}\right)^\theta
\right)\right.\right.\\\nonumber&&\left.\left.
+J^{\theta\theta}-\frac{2b}{B_0}I^{\theta\theta}\right\}\right]+\left(\frac{(cC_0)^\theta}{C_0^2}
-\frac{2b}{B_0}\frac{C_0^\theta}{C_0}\right)
\left[\frac{\alpha R_0^2B_0^2}{2}+I'\left(\frac{A_0'}{A_0}
-\frac{2B_0'}{B_0}+\frac{1}{r}\right)
\right.\\\label{a3}&&\left.+I''+\frac{I^{\theta\theta}}{r^2}-\frac{I^\theta}{r^2}r
\left(\frac{A_0^\theta}{A_0}-\frac{2B_0^\theta}{B_0}\right)\right],
\end{eqnarray}
\begin{eqnarray}\nonumber &&
Z_4=\frac{A_0^2}{2}\left(\frac{B_0C_0}{bC_0-cB_0}\right)\left[\frac{2}{B_0^2}
\left\{\frac{A_0'C_0'}{A_0C_0}\left(\frac{a'}{A'_0}-\frac{a}{A_0}+\frac{c'}{C'_0}
-\frac{b}{B_0}\right)+\frac{A_0''}{A_0}\left(\frac{a''}{A''_0}
\right.\right.\right.\\\nonumber &&\left.\left.\left.-\frac{a}{A_0}\right)
+\frac{B_0''}{B_0}\left(\frac{b''}{B''_0}-\frac{b}{B_0}\right)+\frac{C_0''}{C_0}\left(\frac{c''}{C''_0}
-\frac{b}{B_0}\right)-
\frac{1}{r}\left(\frac{a}{A_0}-\frac{2b}{B_0}\right)'
\right.\right.\\\nonumber &&\left.\left.-\frac{2B_0'}{B_0}\left(\frac{b}{B_0}\right)'+\frac{2}{r^2}\left\{
\frac{2B_0^\theta}{B_0}\left(\frac{b}{B_0}\right)^\theta+
\frac{A_0^{\theta\theta}}{A_0}\left(\frac{a^{\theta\theta}}{A^{\theta\theta}_0}-\frac{a}{A_0}\right)
+\frac{B_0^{\theta\theta}}{B_0}\left(\frac{b^{\theta\theta}}{B^{\theta\theta}_0}
-\frac{b}{B_0}\right)\right.\right.\right.\\\label{a4} &&\left.\left.\left.+\frac{C_0^{\theta\theta}}{C_0}\left(\frac{c^{\theta\theta}}{C^{\theta\theta}_0}
-\frac{b}{B_0}\right)+\frac{A_0^\theta C_0^\theta}{A_0C_0}\left(\frac{a^\theta}{A^\theta_0}-\frac{a}{A_0}+\frac{c^\theta}{C^\theta_0}
-\frac{b}{B_0}\right)\right\}\right\}-e-\frac{2bR_0}{B_0}\right].
\end{eqnarray}

\end{document}